\documentclass{aa} 

\usepackage{graphicx}
\usepackage{txfonts}
\usepackage{natbib}
\bibpunct{(}{)}{;}{a}{}{,}
\usepackage{amssymb}
\usepackage{xfrac}
\usepackage{xcolor}
\usepackage{soul}
\usepackage{ulem}

\newcommand{\figref}[1]{Fig.~\ref{#1}} 
\newcommand{\up}[1]{$^{#1}$}
\newcommand{\down}[1]{$_{#1}$}

\newcommand{\tabref}[1]{Table~\ref{#1}}

\newcommand{\kms}{km~s$^{-1}$}

\newcommand{\degree}{$^{\circ}$}

\newcommand{\dlq}[1]{``}
\newcommand{\drq}[1]{''}
\newcommand{\react}[3]{#1 + #2 $\rightarrow$ #3}
\newcommand{\dec}[4]{#1\degree#2$^\prime$#3\farcs#4}
\newcommand{\ra}[4]{#1$^\mathrm{h}$#2$^\mathrm{m}$#3\farcs#4}

\newcommand{\err}[2]{$^{+#1}_{-#2}$}

\begin{document}

	\title{The ALMA-PILS survey: Formaldehyde deuteration in warm gas on small scales toward IRAS 16293$-$2422 B}

\titlerunning{ALMA-PILS: Formaldehyde deuteration toward IRAS 16293$-$2422 B}

   \author{M.~V. Persson\inst{1,2}
	     \and J.~K. J{\o}rgensen\inst{3}
	     \and H.~S.~P. M{\"u}ller \inst{4}
	     \and A. Coutens\inst{5}
		 \and E.~F. van Dishoeck\inst{2,6}
	     \and V. Taquet\inst{7}
	     \and H. Calcutt\inst{2}
	     \and M.~H.~D. van der Wiel\inst{8}
	     \and T.~L. Bourke\inst{9}
	     \and S. Wampfler\inst{10}
          }

   \institute{
   	Department of Space, Earth and Environment, Chalmers University of Technology, Onsala Space Observatory, 439 92, Onsala, Sweden, \email{magnpe@chalmers.se} 
		   	\and Leiden Observatory, Leiden University, P.O. Box 9513, NL-2300 RA Leiden, The Netherlands
            \and Centre for Star and Planet Formation,
            Niels Bohr Institute \& Natural History Museum of Denmark,
            University of Copenhagen, {\O}ster Voldgade 5--7, DK-1350, Copenhagen K, Denmark
			\and I. Physikalisches Institut, Universit{\"a}t zu K{\"o}ln, Z{\"u}lpicher Str. 77, 50937 K{\"o}ln, Germany	
            \and Laboratoire d'astrophysique de Bordeaux, Univ. Bordeaux, CNRS, B18N, all{\'e}e Geoffroy Saint-Hilaire, 33615 Pessac, France
            \and Max-Planck Institute f\"{u}r extraterrestrische Physik (MPE), 
            Giessenbachstrasse, 85748 Garching, Germany
            \and INAF-Osservatorio Astrofisico di Arcetri, Largo E. Fermi 5, I-50125 Firenze, Italy	    
            \and ASTRON Netherlands Institute for Radio Astronomy, PO~Box~2, 7990 AA~Dwingeloo, The Netherlands
		    \and SKA Organization, Jodrell Bank Observatory, Lower Withington, Macclesfield, Cheshire SK11~9DL, UK
			\and Center for Space and Habitability (CSH) University of Bern Sidlerstrasse 5, CH-3012 Bern, Switzerland
                }

   \date{Received XX July, 2017; accepted XX YY, 201Z}

 
  \abstract
   {
   	The enhanced degrees of deuterium fractionation observed in envelopes around protostars demonstrate the importance of chemistry at low temperatures, relevant in pre- and protostellar cores.
   	Formaldehyde is an important species in the formation of methanol and more complex molecules. 
   	}
   {
	Here, we present the first study of formaldehyde deuteration on small scales around the prototypical low-mass protostar IRAS~16293$-$2422 using high spatial and spectral resolution Atacama Large Millimeter/submillimeter Array (ALMA) observations. We determine the excitation temperature, abundances and fractionation level of several formaldehyde isotopologues, including its deuterated forms.
   	}
   {
   	Excitation temperature and column densities of formaldehyde in the gas close to one of the components of the binary are constrained through modeling of optically thin lines assuming local thermodynamical equilibrium. The abundance ratios are compared to results from previous single dish observations, astrochemical models and local ISM values.
   	}
   {
   	Numerous isotopologues of formaldehyde are detected, among them H$_2$C$^{17}$O, and D$_2^{13}$CO for the first time in the ISM. The large range of upper energy levels covered by the HDCO lines help constrain the excitation temperature to 106$\pm$13~K. 
   	Using the derived column densities, formaldehyde shows a deuterium fractionation of HDCO/H$_2$CO=6.5$\pm$1\%, D$_2$CO/HDCO=12.8\err{3.3}{4.1}\%, and D$_2$CO/H$_2$CO=0.6(4)$\pm$0.1\%.
   	The isotopic ratios derived are \up{16}O/\up{18}O=805\err{43}{79}, \up{18}O/\up{17}O=3.2\err{0.2}{0.3} and \up{12}C/\up{13}C=56\err{8}{11}{$\quad$}.
   	}
   {
   	The HDCO/H\down{2}CO ratio is lower than found in previous studies, highlighting the uncertainties involved in interpreting single dish observations of the inner warm regions. The D\down{2}CO/HDCO ratio is only slightly larger than the HDCO/H\down{2}CO ratio. This is consistent with formaldehyde forming in the ice as soon as CO has frozen onto the grains, with most of the deuteration happening towards the end of the prestellar core phase. A comparison with available time-dependent chemical models indicates that the source is in the early Class~0 stage.
   }


   \maketitle
%

\section{Introduction}
Among molecular abundance ratios, the deuterium fractionation (D/H ratio) is commonly used to infer the formation conditions \citep[e.g.][]{roberts00b}. In general, a high D/H ratio indicates a low temperature, and a low D/H ratio indicates a high temperature during formation. The deuterium-to-hydrogen ratio in the local interstellar medium (ISM) is 2.0$\pm$0.1$\times$10\up{-5} \citep{prodanovic10}. Many molecules are formed on the surfaces of dust grains, and once it is cold enough for CO to freeze out onto the grains, the deuterium chemistry is enhanced. This means that any molecules formed after CO freeze-out are expected to have high levels of deuterium fractionation \citep[e.g.][]{taquet14}. 
Single dish observations of various molecules toward deeply embedded protostars have shown high deuterium fractions, above 10\%\ \citep[e.g.][]{vandishoeck95,parise06}. Optical depth and source size effects (beam dilution) are significant for single dish observations of lines of the main isotopologue which makes the uncertainty in the fractions large. High sensitivity and high resolution interferometric observations have the possibility to circumvent these problems since they probe smaller scales, close to the forming star where the temperature is high enough to sublimate the grain surface ices completely (i.e.\ $T\gtrsim100$~K). The higher sensitivity makes it possible to detect weaker lines of minor isotopologues where optical depth effects are less important.

Observations of water have revealed a low deuterium fractionation (e.g.\ HDO/H\down{2}O ratio) on small scales (warm gas) in young protostellar envelopes. While the cold gas shows a HDO/H\down{2}O ratio of a few \%, the warm gas has a ratio closer to 0.1\% \citep[e.g.][]{persson14,coutens12}. Furthermore, the water deuterium fractionation of the warm gas in the inner region toward the protostar NGC~1333 IRAS~2A shows a D\down{2}O/HDO ratio significantly higher than the HDO/H\down{2}O ratio \citep[seven times higher,][]{coutens14}. \citet{furuya16} explain the observed ratios by modeling the pre-stellar core physical and chemical evolution. 
The gas-phase deuteration processes are inefficient early on due to the higher H\down{2} o/p-ratio raising the destruction rate of the main seed molecule for deuteration processes, H$_2$D$^+$ \citep{pagani92}. However, the deuteration processes through H$_2$D$^+$ increase as soon as the o/p-ratio decreases and other molecules that destroy it freezes out \citep[e.g.\ CO,][]{furuya15}.
This and the low temperature results in higher gas-phase atomic D/H ratio which in turn causes more D atoms to accrete (mainly through H$_2$D$^+$) on the grains \citep[first pointed out by][]{tielens83}. This drives the hydrogen-surface deuterium chemistry, however the total production rate of water is significantly decreased at this point, thus for water the deuteration is low. This explains both the low absolute deuteration that water has, and the relative ratios between the deuterated forms. Whether other molecules show similar trends is not yet clear. However many molecules are thought to form later, once CO has frozen onto the grains and could thus show different ratios in this model.

Formaldehyde is an important molecule, not only for constraining physical conditions in the gas of star-forming regions, but also as an intermediary in the formation path to complex organic molecules. 
H\down{2}CO formation in ices was shown experimentally and models highlighted that it proceeds at a significantly higher rate in ices than in gas-phase reactions \citep{roberts04}.
Thus it is thought that formaldehyde is mainly formed on the surface of dust grains once CO has frozen onto the grains through the hydrogen addition reactions \react{CO}{H}{HCO} and then \react{HCO}{H}{H$_2$CO} seen in the laboratory \citep{watanabe02,fuchs09}. The deuterated forms of H$_2$CO can be obtained through subsequent substitution reactions involving D-atoms, i.e. \react{H\down{2}CO}{D}{HDCO+H} and \react{HDCO}{D}{D\down{2}CO+H}. They can also be obtained through abstraction and addition reactions through the HCO and DCO radical \citep[e.g.][]{tielens83}, which has been shown to be efficient at low temperature in ices \citep{hidaka09}. Naturally, just as with H\down{2}CO there is also a direct channel from CO through DCO, i.e. the tunneling reaction \react{CO}{D}{DCO}, with subsequent addition reactions to form D\down{2}CO and HDCO. 

Assuming a grain surface formation route, \citet{turner90} and later \citet{charnley97} estimated the relation between abundance ratios of the deuterated isotopologues of formaldehyde, assuming equal transmission probabilities for the CO+H and CO+D reaction channels and neglecting abstraction reactions, to be
\begin{equation}\label{eq:1}
\dfrac{\rm D_2CO}{\rm HDCO} = \dfrac{1}{4} \dfrac{\rm HDCO}{\rm H_2CO} .
\end{equation}
While this was enough to explain the observed deuterium fractionation of formaldehyde at the time, it seems the situation is slightly more complex for deeply-embedded protostars, and that substitution and abstraction (with subsequent addition) reactions are important.

In this paper we study the deuterium fractionation of formaldehyde (H$_2$CO) on small solar-system scales toward one of the sources in the Class~0 multiple system IRAS~16293$-$2422, located in the $\rho$~Ophiuchus star forming region at a distance of 120~pc \citep{loinard08}. The deuterium fractionation is determined through line analysis of various deuterated and non-deuterated isotopologues of formaldehyde. IRAS~16293$-$2422 is a protostellar binary with a separation of about 5\arcsec\ (600~AU), where the SE source is referred to as "A" and the NW source as "B". Several studies have shown IRAS~16293$-$2422 to be a chemically rich source, with a wealth of complex organic molecular species \citep[e.g.][]{bottinelli04,kuan04,bisschop08,jorgensen11} and with thermal water \citep{persson13} associated with both sources. Recent detections include the prebiotic molecule glycolaldehyde \citep{jorgensen12}, its deuteration \citep{jorgensen16a}, and the detection of ethylene oxide, acetone, propanal, methyl isocyanate and formamide \citep{coutens16,lykke17,martin17,ligterink17}. For an extended overview of the literature on IRAS~16293$-$2422 and a review of the ALMA-PILS survey see \citet{jorgensen16a}.

Measurements of the deuterium fractionation of formaldehyde toward IRAS~16293$-$2422 have previously been limited to single dish observations. Using the JCMT and the CSO \citet{vandishoeck95} derived a HDCO/H\down{2}CO ratio of 14$\pm$7\% in a 20\arcsec\ beam ($2400$~AU), taking non-LTE effects into account. Assuming the same excitation temperature for both HDCO and H\down{2}CO an even higher ratio of HDCO/H\down{2}CO=33\err{42}{18}\% was obtained with the same observations by \citet{parise06}. The D\down{2}CO/H\down{2}CO ratio was mapped around the source and constrained to 3\% toward the edge of the envelope and peaking at 16\% toward a position roughly one single dish beam south \citep{ceccarelli98,ceccarelli01}. Further single dish observations, also with signatures of self-absorption, constrain the ratios to HDCO/H\down{2}CO=13--16\%, D\down{2}CO/H\down{2}CO = 5--6\%, and D\down{2}CO/HDCO= 33--40\% \citep{loinard00}. However, for these studies opacity effects and multiple contributions to the emission within the beam might affect the column densities by factors of a couple to a few \citep{parise06}, thus the abundances in cold gas need to be more accurately constrained for a proper comparison with ratios derived here for the warm gas on small scales. 

Formaldehyde was mapped at high angular resolution toward IRAS~16293$-$2422 by \citet{schoier04}, showing some of the uncertainties involved in deriving abundances of formaldehyde with single dish telescopes, and highlighting the importance of higher resolution. 

The formation and deuteration of formaldehyde, methanol and other species were studied through chemical modeling by \citet{taquet12} and more recently by \citet{taquet14}. The more recent models trace the deuteration with time from the prestellar core stage to the end of the deeply-embedded stage. The models cannot reproduce the high levels of formaldehyde deuteration previously observed. To further test the chemical models and assess the importance of grain surface formation as well as the various formation paths it is important to accurately constrain the relative abundances of the different deuterated forms of formaldehyde. Interferometric observations of the inner warm region have the possibility to constrain the abundances where the grain surface ice has just been completely sublimated, without the pitfalls of single dish observations.

This paper is laid out as follows. In Sect.~\ref{sect:obs} details of the observations and spectroscopic data are given together with a description of the analysis method. Section~\ref{sect:results} gives the result of the analysis; the measured excitation temperature and the column densities of the various isotopologues of formaldehyde. In Section~\ref{sect:discussion} the derived abundances and isotopologue ratios are discussed with previous measurements and chemical models in mind. This is followed by a short summary and outlook on future prospects in Sect.~\ref{sect:future}.

%

\section{Observations and Analysis}\label{sect:obs}

This study is based on observations from the Protostellar Interferometric Line Survey (PILS\footnote{\url{http://youngstars.nbi.dk/PILS}}), an ALMA band 7 unbiased spectral line survey with complete coverage between 329.15 and 362.90~GHz at 0.244~MHz resolution. For details, including observing conditions, calibration and imaging, see \citet{jorgensen16a}. The observations include both the main array of 12 m dishes, and the 7 m dishes of the Atacama Compact Array (ACA). The observations cover both sources in the IRAS~16293$-$2422 multiple system. The PILS spectrum toward source~A shows lines with larger line widths (2--8~\kms\ FWHM) than for source~B (around 1~\kms\ FWHM) making line identification much easier toward the B~source, also when comparing to the Galactic Center and high-mass protostars previously observed in the same fashion. The inclusion of the ACA observations fills in the shorter baselines and thus the extended emission encapsulating the sources with a maximum recoverable scale of $\sim$13\arcsec. The final spectral image cubes have a RMS of 4--8~mJy~beam$^{-1}$~km~s$^{-1}$, a synthesized beam with a FWHM of 0\farcs5 and a flux calibration uncertainty of $\sim5\%$. The phase center is located half-way between the large scale binary A and B ($\alpha_\mathrm{J2000}$ = \ra{16}{32}{22}{72}; $\delta_\mathrm{J2000}$ = \dec{$-$24}{28}{34}{3}) and the field of view covers roughly 15\arcsec. 

The observed spectra toward the two main sources show a large number of lines. The unprecedented sensitivity and richness of the observed spectrum have so far facilitated several new detections and constraints on isotopologue ratios. So far the abundances and excitation conditions for acetone, propanal and methyl isocyanate \citep{lykke17,ligterink17}, and also the deuterium fraction of formamide, glycolaldehyde, ketene and other oxygen-bearing (complex) organic molecules have been presented \citep{coutens16,jorgensen16a,jorgensen17}.

\begin{figure}[ht]
	\centering
	\includegraphics[width=\linewidth]{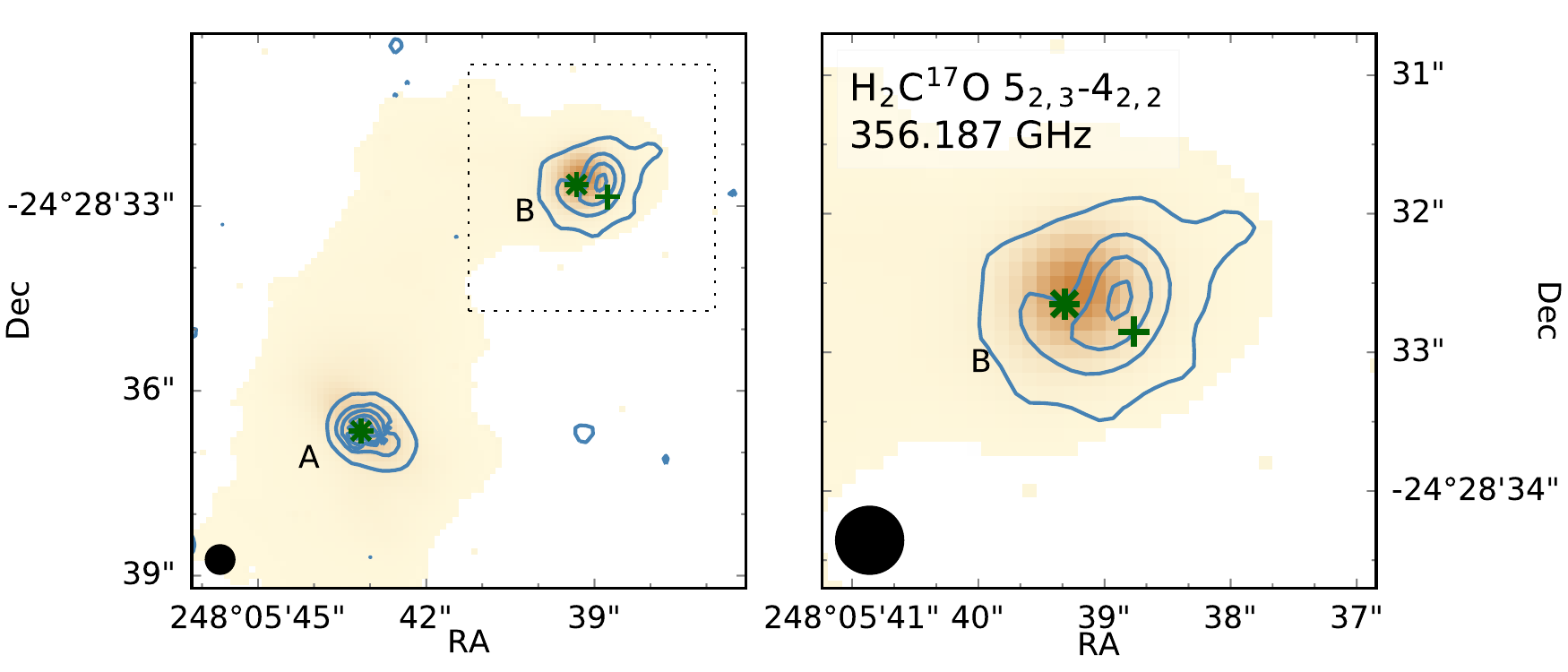}
	\caption{PILS ALMA continuum map of the IRAS~16293$-$2422 system at 850~${\mu}m$. The background image shows the continuum emission, the contours indicate the H$_2$C$^{17}$O 5$_{2,3}$--4$_{2,2}$ line (E$_u$=98.5~K) integrated emission, the star is the position of the continuum peak, and the plus sign the position of the extracted spectrum (one beam, i.e.\ 0\farcs{5}, away from the continuum). The continuum image is cut off at 15~mJy ($3\sigma$) and the line contours start at 25~mJy~km~s\up{-1} ($5\sigma$) in steps of 120~mJy~km~s\up{-1}.}
	\label{fig:extract_position}
\end{figure}

Similar to the analysis presented in \citet{lykke17}, \citet{coutens16} and \citet{ligterink17}, the spectrum from an offset position toward source~B was extracted. Located at $\alpha_\mathrm{J2000}$=\ra{16}{32}{22}{58} $\delta_\mathrm{J2000}$=\dec{$-$24}{28}{32}{8} it is $\sim$0\farcs5 offset (i.e.\ one beam) in the Southwest direction relative to the continuum peak of source~B (see \figref{fig:extract_position}). The main reason for this is that toward the continuum peak, lines are affected by varying degrees of absorption and continuum optical depth, complicating the line identification and analysis of the abundances (see contours in the right panel of \figref{fig:extract_position}).

\subsection{Laboratory spectroscopy data}\label{sect:H2CO-lab-spec}

The transition frequencies and other data (including partition function values) of all forms of formaldehyde were taken from the CDMS database \citep{muller01,muller05}. 

The H$_2$CO, H$_2$C$^{17}$O and H$_2$C$^{18}$O entries are based on \citet{muller17}, and the H$_2^{13}$CO entry on \citet{muller00c}. Important additional data are from \citet{brunken03,bocquet96} in the case of H$_2$CO along with ground state combination differences employed in \citet{muller00a}. \citet{muller00b,cornet80a} contributed additional data to the H$_2$C$^{18}$O entry, the latter also to those of H$_2$CO and H$_2^{13}$CO. The H$_2$C$^{17}$O entry employed previous data from \citet{cornet80b} and from \citet{flygare65}. 

Entries of the deuterated isotopologues, HDCO, D$_2$CO, HD$^{13}$CO, D$_2^{13}$CO, HDC$^{18}$O and D$_2$C$^{18}$O, were based on the very recent analyses by \citet{zakharenko15}. The analyses took into account earlier data for five of these six isotopologues (no earlier data for HDC$^{18}$O) from \citet{dangiosse78} as well as extensive data  for HDCO and D$_2$CO from \citet{bocquet99}. Far-infrared data of D$_2$CO \citep{lohilahti04} and D$_2^{13}$CO \citep{lohilahti05} were also employed for the CDMS entries.


\subsection{Abundances}
\begin{figure}[hb]
	\centering
	\includegraphics[width=\linewidth]{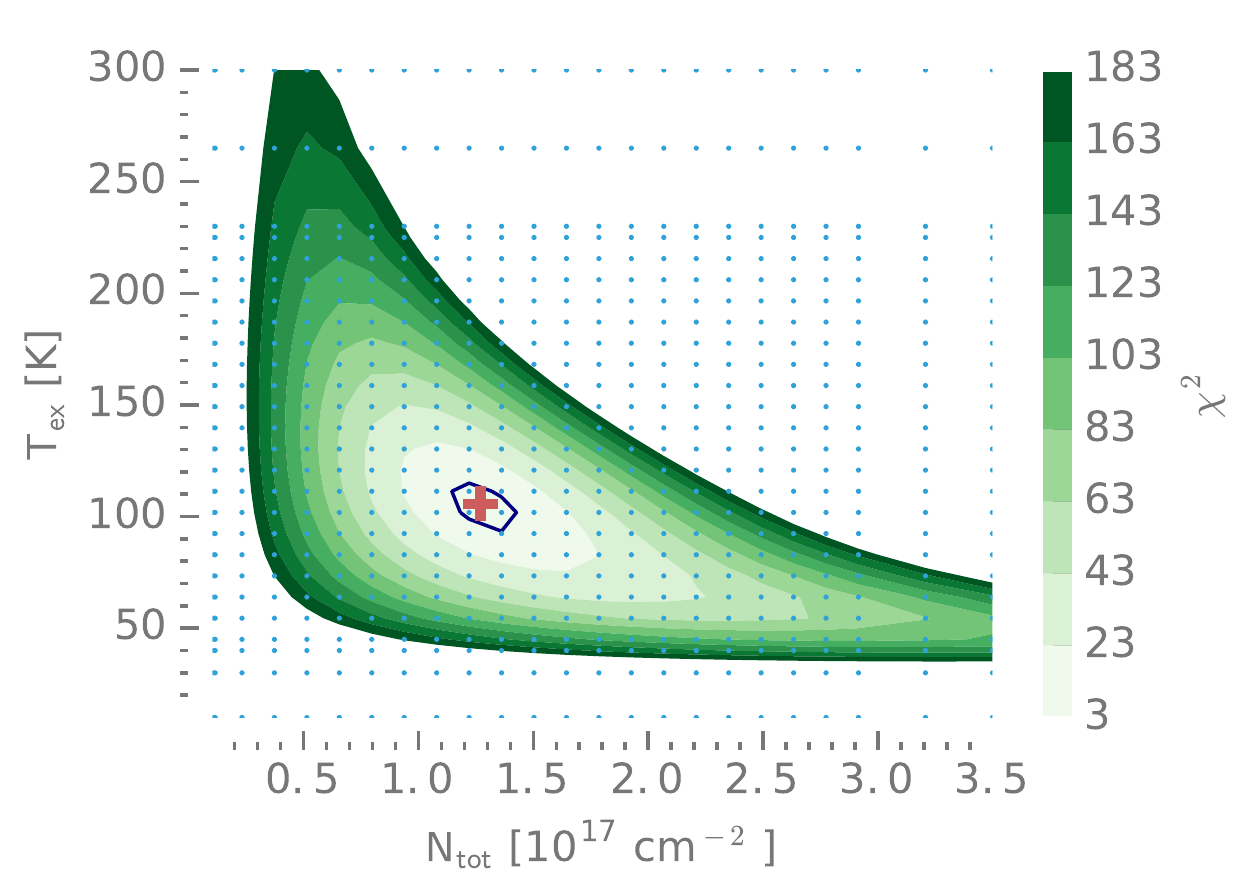}
	\caption{Grid constraining the excitation temperature and column density of HDCO. The cross marks the best fit values, $1.3\pm0.2\times10^{17}$~cm\up{-2} and $106\pm13$~K from interpolation of the values within the marked area. The contour line marks the $\Delta\chi^2=2.3$ region. Note that the column density value has been corrected for the continuum. The dots show the evaluated column densities and excitation temperatures.}
	\label{fig:hdco_tex_ntot_grid}
\end{figure}

The abundances are estimated by computing a synthetic spectrum assuming Local Thermal Equilibrium (LTE) and comparing it with the observed spectrum. In addition to this, CASSIS\footnote{Developed by IRAP-UPS/CNRS \url{http://cassis.irap.omp.eu}} was used as a support tool to e.g.\ check for possible line blending from other species, which is the same method as in other PILS survey studies e.g.\ \citet{coutens16, ligterink17,lykke17}. For HDCO a full grid in excitation temperature and column density is calculated and the coordinates of the minimum $\chi^2$ are taken as the best estimates. Because the excitation temperature is better constrained using HDCO than through the other isotopologues of formaldehyde, its resulting temperature is used to constrain the column density for all forms. This is done by varying the column density, calculating the integrated line flux for each line not affected by optical depth effects or significant blending by other species and comparing to the observed line flux by calculating the $\chi^2$ estimate. The uncertainty in abundance is estimated by finding the minimum $\chi^2$ for the lower and upper bound excitation temperature (i.e.\ 93 and 119~K, \figref{fig:hdco_tex_ntot_grid}). The column densities are corrected for the continuum emission from the surrounding dense dust \citep[by multiplying with a factor of 1.1658,][]{jorgensen16a}. A line width of 1.15~km~s\up{-1} (FWHM) and system velocity of 2.7~km~s\up{-1}\ were used (relevant for source B). The assumed line width agrees well with the observed line profiles for formaldehyde and is similar to that of other species \citep{coutens16,lykke17}.

The detected transitions of the various isotopologues, deuterated and non-deuterated are listed in Table~\ref{tab:lines} in the Appendix. The analysis focuses on H\down{2}CO (3 lines detected) and several deuterated and non-deuterated isotopologues (number of detected lines in parenthesis): H\down{2}\up{13}CO (9), H\down{2}C\up{18}O (9), H\down{2}C\up{17}O (8), HDCO (6), HDC\up{18}O (1 tentative), D\down{2}CO (11), and D\down{2}\up{13}CO (10). 

%

\section{Results}\label{sect:results}
In the following sections we present the results of a detailed analysis of the column densities, the various isotopic ratios and the constraints on the excitation temperature. Only optically thin lines are included in the analysis to minimize any opacity effects.

\subsection{Excitation temperature}\label{sect:tex}
The excitation temperature was determined to be $106\pm13$~K by fitting the spectrum of HDCO, and the resulting temperature is used for all forms of formaldehyde to constrain their column density. The $\chi^2$ in a grid around the best fitting column density and excitation temperature is shown in \figref{fig:hdco_tex_ntot_grid}. The region with an added $\Delta\chi^2=2.3$ is taken as the $1\sigma$ error in $T_\mathrm{ex}$ and $N_\mathrm{tot}$ for HDCO \citep{avni76,press02}. This temperature, together with the constrained column densities reproduce the lines for all isotopologues well. The figures in Appendix~\ref{app:figs_lines} show the LTE model for the best fitting column densities and the observed spectrum for all lines of all species, including non-detections.

In some studies of other sources  a second temperature component is needed to reproduce the spectra of lines of both lower and higher upper energy levels \citep[e.g.][]{isokoski13}. If a second temperature component were relevant for this study, e.g.\ the low E$_u$ lines would be systematically under/over-produced in the synthetic spectrum. Since this is not the case it is likely that only one temperature component is present. The other isotopologues were not as suitable to use for constraining the excitation temperature, the resulting parameter space of HDCO shows a clearer minimum than any of the other isotopologues investigated (\figref{fig:hdco_tex_ntot_grid}). The spread in E$_u$ of the detected and optically thin lines of HDCO makes it better at constraining the temperature. Using selected lines of H\down{2}C\up{18}O a less constrained excitation temperature is obtained ($90\sim200$~K). For the other isotopologues, it was possible to constrain the column density to high accuracy with the given excitation temperature.

In addition to the detected, optically thin HDCO lines (\tabref{tab:lines}) we also used upper limits of one non-detected HDCO transition at 348.965~GHz ($E_u=1017.78$~K, $\log_{10}(A_{ij})=-4.5321$) to constrain the excitation temperature (see \tabref{tab:lines_ulims}). Since the LTE model should not overproduce this line, it increases the constraint. Thus, a wide spread in $E_u$ was used to constrain the $T_\mathrm{ex}$, but avoiding the $11_{1,10}-11_{1,11}$ line at 346.74~GHz which is optically thick. 

\subsection{Column densities and ratios}
The best fit column densities at $T_\mathrm{ex}=106\pm13$~K for the species are listed in \tabref{tab:ntots}. The uncertainty in abundance is dominated by the uncertainty in excitation temperature. In \tabref{tab:ratios} the various column density ratios for the studied isotopologues are shown. Ratios where multiple estimates exist using isotopologues, i.e.\ HDCO / H\down{2}CO and HDC\up{18}O/H\down{2}C\up{18}O for the HDCO/H\down{2}CO ratio, are combined. When averaging/combining values with asymmetric uncertainties care has to be taken when calculating the best combined estimate \citep[for a detailed discussion and solution see Section 3 in ][]{barlow03}. To constrain the column density of the main isotopologue H\down{2}CO only one detection and one upper-limit (non-detection) are used in the fit. The other H\down{2}CO transitions are optically thick, by a large margin, and they also show some extended emission (van der Wiel, in prep.).
\begin{table}
	\caption{Best-fit continuum corrected column densities at $T_\mathrm{ex}=106\pm13$~K for all the isotopologues of formaldehyde.}\label{tab:ntots}
	\centering
	\begin{tabular}{lp{60pt}}
		\hline\hline
		Species & $N_\mathrm{tot}$ (cm$^{-2}$) \\
		\hline
		H\down{2}CO 		& 1.9\err{0.1}{0.2}\,$\times$10$^{18}$ \\
		H\down{2}\up{13}CO 	& 3.6\err{0.7}{1.1}\,$\times$10$^{16}$ \\
		H\down{2}C\up{17}O 	& 7.2\err{0.5}{0.6}\,$\times$10$^{14}$ \\
		H\down{2}C\up{18}O 	& 2.3$\pm0.1$\,$\times$10$^{15}$ \\
		HDCO 				& 1.3$\pm0.2\,\times$10$^{17}$ \\
		HDC\up{18}O 		& $\leq$1.4$\pm0.4\,\times$10$^{14}$ \\
		D\down{2}CO 		& 1.6\err{0.4}{0.5}\,$\times$10$^{16}$ \\
		D\down{2}\up{13}CO 	& 2.2\err{0.2}{0.4}\,$\times$10$^{14}$ \\
\hline
	\end{tabular}
\end{table}

\begin{table}
	\caption{Derived fractionation ratios for the various forms of formaldehyde. The errors are propagated from the uncertainty in the determined column densities.}\label{tab:ratios}
	\centering
	\begin{tabular}{llll}
		\hline\hline
		 Species 							& Ratio 				& Comb.\tablefootmark{a} & D/H\\
		\hline
		HDCO / H\down{2}CO 					& 6.8\err{1.1}{1.3}\,\% & 6.5$\pm$1\,\% & 3.25$\pm$1\,\% \\
		HDC\up{18}O/H\down{2}C\up{18}O 		& 6.0$\pm1.5$\,\%  		&  --			& \\
		D\down{2}CO / HDCO 					& 12.8\err{3.3}{4.1}\,\%&  				& 25.6\err{3.3}{4.1}\,\% \\
		D\down{2}CO / H\down{2}CO 			& 0.9\err{0.2}{0.3}\,\% & 0.6(4)$\pm$0.1\,\% & 8.0$\pm$0.1\% \\
		D\down{2}\up{13}CO/H\down{2}\up{13}CO& 0.6\err{0.1}{0.2}\,\%&  --			& \\
		\hline
		H\down{2}CO / H\down{2}\up{13}CO 	&  52\err{10}{16}  & 56\err{8}{11} & \\
		 D\down{2}CO / D\down{2}\up{13}CO 	&  74\err{18}{24}  & --  & \\
		\hline
		 H\down{2}CO / H\down{2}C\up{18}O 	& 800\err{47}{98}  & 805\err{43}{79} & \\
		HDCO / HDC\up{18}O 					& 908$\pm263$      & -- & \\
		\hline
		 H\down{2}CO / H\down{2}C\up{17}O 	& 2596\err{639}{395}  &  & \\
		\hline
     	 H\down{2}C\up{18}O / H\down{2}C\up{17}O & 3.2\err{0.2}{0.3}  & &  \\
	    \hline
	\end{tabular}
	\tablefoot{
	\tablefoottext{a}{Combined isotopologue ratio \citep[following][]{barlow03}.}	
	}
\end{table}

\section{Discussion}\label{sect:discussion}

\subsection{Column densities and excitation temperature}\label{dis:texandntot}

The derived excitation temperature of 106$\pm$13~K agrees with the results for some of the other molecules studied within the PILS survey. Acetaldehyde, ketene, ethylene oxide, acetone, propanal, and also dimethyl ether show a similar excitation temperature, on the order of 125~K \citep{lykke17,jorgensen17}. Interestingly \citet{vandishoeck95} derived, after subtracting the extended contribution from the central emission, a kinetic temperature of 100\err{40}{20}~K from the H\down{2}CO 3\down{22}--3\down{21}/3\down{03}--2\down{02} ratio toward IRAS~16293$-$2422.

The derived column densities fit the observed spectrum well using the assumed $T_\mathrm{ex}$ of 106~K from HDCO. The H\down{2}\up{13}CO line at 355.04~GHz is partly blended with the weaker counterpart of H\down{2}C\up{17}O. This means that the H\down{2}\up{13}CO column density might be overestimated by a small amount, however the effect is within the uncertainties. A H\down{2}C\up{17}O column density and excitation temperature taken at the lower uncertainty limit will make its blending effect on H\down{2}\up{13}CO insignificant. 

\subsection{Deuterium fractionation}
In this section we present the deuterium fractionation calculated from the observations and then compare these to the model of \citet{taquet14}.

\subsubsection{Measured D/H ratios}
Together with the column density ratios for the studied isotopologues in \tabref{tab:ratios} are the best estimated D/H ratios as well (combined measurements where applicable), where D/H is given by HDCO/H\down{2}CO$\times0.5$, D\down{2}CO/HDCO$\times$2 and $\sqrt{\mathrm{D_2CO/H_2CO}}$. While the errors quoted for the column densities are derived by fitting spectra with the column density for the highest and lowest excitation temperature, the errors in the ratios were derived with error propagation. Thus the uncertainties of the column densities and in extension the fractionation ratios, reflect also the uncertainty in excitation temperature.

The HDC\up{18}O/H\down{2}C\up{18}O ratio is 6.0\% (D/H ratio of 3.0\%), and it agrees with the main isotopologues HDCO/H\down{2}CO ratio of 6.8\% (D/H ratio of 3.4\%), confirming the method used for constraining the column densities. Their combined ratio 6.5\% (D/H ratio 3.25\%) is taken as the best estimate. The D\down{2}\up{13}CO/H\down{2}\up{13}CO ratio (0.6\%) is also similar to the main form D\down{2}CO/H\down{2}CO ratio (0.9\%), their combined ratio being 0.6(4)\% (D/H ratio 8.0\%). Finally, the D\down{2}CO/HDCO ratio is 12.8\% (D/H ratio is 25.6\%).

These numbers show that the D\down{2}CO/HDCO ratio is two times higher than the HDCO/H\down{2}CO ratio, significantly more than the \sfrac{1}{4} expected from statistical arguments (see Equation~\ref{eq:1} and related text). Furthermore, other organics show D/H ratios similar to this \citep{jorgensen17}. This shows that for a full understanding a more complex interpretation than simple statistical arguments is needed. Whether other protostars show similar deuterium fractionation ratios for H\down{2}CO as deduced here is not yet clear, observations of more protostars are needed to investigate what role initial environment and evolution might play in determining the ratios.

\subsubsection{Models}
\citet{taquet12} and more recently \citet{taquet14} studied the evolution of deuterated ices during the earliest stages of star formation, including the prestellar and protostellar (Class~0) phase. In the \citeyear{taquet12} study the importance of abstraction and substitution reactions is shown, as also supported by lab data \citep{hidaka09}. Furthermore, the HDCO/H\down{2}CO ratio is not affected by the inclusion of abstraction reactions due to the efficiency of the deuterium abstraction on HDCO, forming back H\down{2}CO. Thus, constraining the deuterium fractionation in all forms of formaldehyde, especially the doubly deuterated form is crucial for constraining the formation conditions and the chemical models. The results show that a reaction network without abstraction and substitution reactions cannot reproduce the HDCO/H\down{2}CO and D\down{2}CO/HDCO ratios simultaneously. Including abstraction reactions shows that the ratios are reproducible at moderate densities ($n_\mathrm{H}=10^6$~cm$^{-3}$). The model in the \citeyear{taquet14} study traces the density and temperature evolution from a (inside-out) collapsing core to the end of the deeply-embedded stage (Class~0). The observed (combined, best estimate) ratios of HDCO/H\down{2}CO of 6.5$\pm$1\% and D\down{2}CO/H\down{2}CO of 0.6(4)$\pm$0.1\% are best reproduced by the model at $t=1.1\times10^5$~years. This timescale represents the beginning of the Class~0 stage, where the D/H ratio in HDCO is 4.1\%\ and 0.18\%\ for D\down{2}CO (Table~7 in \citeauthor{taquet14}). While the density and temperature profile at this time step from \citeauthor{taquet14} is similar to the physical model of \citet{crimier10} for IRAS~16293-2422 specifically, the modeled deuterium fractionation observed in the hot corino gas is mostly due to the deuteration processses of formaldehyde in ices in the previous dense cloud stage and thus less dependent on the current physical structure. In \citeauthor{taquet14} the assumed size of the beam in the observations of warm gas with interferometers is 0\farcs4 (or 50~AU) which is very similar to the beam in this study (0\farcs5). The excitation temperature derived shows that we are probing the warm gas of the hot corino.
The ratios derived using single dish telescopes, which may apply to larger scales, are more difficult to compare with the model results. It is not clear to which extent opacity effects, assumed size of emitting region and contributions of various emission components play a role in the previous single dish observations, thus we refrain from any extensive comparison with those results. 

The main production phase of deuterated molecules comes towards the end of the prestellar core stage \citep{furuya16}. This is in agreement with previous studies of starless cores using N\down{2}H\up{+} and N\down{2}D\up{+} \citep{crapsi05}. \citet{taquet14} also trace the deuteration through these early stages, with similar conclusions. The model reproduces the large differences between singly and doubly deuterated forms with the D\down{2}O/HDO ratio being about $7\times$ higher than the HDO/H\down{2}O ratio. In formaldehyde toward IRAS~16293$-$2422 the D\down{2}CO/HDCO ratio is about $2\times$HDCO/H\down{2}CO, a similar but much less pronounced effect than for water. 

\subsubsection{Formaldehyde and methanol}
Methanol (CH\down{3}OH) was studied in the framework of the ALMA-PILS survey \citep{jorgensen17}. The deuterium fractionation (D/H) derived toward IRAS~16293$-$2422~B in methanol is 2\%. This is comparable to singly deuterated formaldehyde. Interesting enough, the time scale of the astrochemical model presented by \citet{taquet14} that best reproduces this level of fractionation in methanol is also $t=1.1\times10^5$~years. This could indicate that IRAS~16293$-$2422~B is indeed a very young Class~0 source. Using the column density derived for methanol by \citeauthor{jorgensen17} the H\down{2}CO/CH\down{3}OH ratio is 0.19 toward IRAS 16293$-$2422~B. A ratio of 1 was inferred toward the inner regions of the same source by \citet{maret05} using single dish observations and radiative transfer modeling, clearly different from the ratio presented in this study. However, the large uncertainties inherent in using single dish measurements to constrain inner envelope abundances makes a comparison difficult. The measured H\down{2}CO/CH\down{3}OH ratio of 0.19 is close to what has been constrained in ices toward protostars, ranging from 0.1 to 0.67 \citep{keane01}. This strengthens the view that warm gas in the innermost regions of protostars represent the bulk ice composition. The measured H\down{2}CO/CH\down{3}OH ratio is in agreement with what was measured in hot cores by \citet{bisschop07}, where the mean and $1\sigma$ standard deviation presented toward seven hot cores was $0.22\pm0.05$.

\subsection{Other isotopic ratios}

While the \up{16}O/\up{18}O ratio is 500 at this distance from the Galactic center \citep{wilson94,wilson99}, the ratio measured in formaldehyde for IRAS 16293$-$2422~B is 800 for H\down{2}CO and 908 for HDCO, both slightly higher than expected at these distances from the Galactic center. The combined ratio for the two measurements is 805\err{43}{79}. Given the uncertainties involved, including the Galactic gradient, \up{16}O/\up{18}O=400--600 at 8~kpc, it is difficult to draw any conclusions, although the combined best estimate ratio is higher than the range in the local ISM.

The \up{18}O/\up{17}O ratio in formaldehyde toward IRAS 16293$-$2422~B is 3.2\err{0.2}{0.3} (H\down{2}C\up{18}O/H\down{2}C\up{17}O). Upper limits of the isotopic variant HDC\up{17}O derived using the $6_{1,6}-5_{1,5}$ transition at 360.962~GHz gives a \up{18}O/\up{17}O ratio of $\geq2$, compatible with the local ISM value and H\down{2}C\up{18}O/H\down{2}C\up{17}O ratios. \citet{jorgensen02} measured the C\up{18}O/C\up{17}O ratio toward the envelopes of 19 protostars, including Class~0 and I sources. Toward IRAS~16293$-$2422 the measured ratio was 3.9, in agreement with the ratio measured for formaldehyde, and roughly 74\% of all the measured sources showed a ratio 2--4. The H\down{2}C\up{18}O/H\down{2}C\up{17}O ratio was first measured toward the Galactic center, 4~kpc molecular ring and the local ISM to be $3.2\pm0.2$ \citep{penzias81}. \citet{wouterloot08} discovered a tentative gradient in the  \up{18}O/\up{17}O ratio with galactocentric distance ranging from 3 in the center to 5 in the outer galaxy, and more recently \citet[][e.g.\ Fig.~3]{li16} strengthened this conclusion. The ratio in the outermost galaxy, beyond 11~kpc is still uncertain. The observed gradient is consistent with an inside-out formation of the galaxy. The observed value toward IRAS~16293$-$2422 fits in the galactic \up{18}O/\up{17}O gradient and could indicate a fairly normal formation environment at this distance from the Galactic center. The deviating \up{16}O/\up{18}O ratio (see previous paragraph) is not fully consistent with this conclusion.

The \up{12}C/\up{13}C ratio for H\down{2}CO and D\down{2}CO and their combined ratio of 56\err{8}{11} are consistent with the ratio of 68$\pm30$, which is relevant for these galactocentric distances \citep{wilson94,wilson99,milam05}. While lines of the isotopic variant HD\up{13}CO are too weak to detect in our survey, its upper limit is compatible with a \up{12}C/\up{13}C ratio of $\geq55$ (using mainly the $11_{1,10}-11_{1,11}$ transition). Recent results from the ALMA-PILS survey presented in \citet{jorgensen17} show that the \up{12}C/\up{13}C ratio in glycolaldehyde, ethanol, methyl formate, and dimethyl ether is around 20--40, lower than for formaldehyde, methanol and the surrounding ISM. This is attributed to a later formation of the low \up{12}C/\up{13}C ratio species, when the availability of \up{13}C is higher.

\subsection{The path to more complex molecules}

Since formaldehyde is thought to play a role in the formation of more complex molecules, it is interesting to compare the deuteration between formaldehyde and its potential daughter species. 
Formaldehyde may also lead to the formation of formamide (NH$_2$CHO) on grain surfaces as shown by some laboratory experiments \citep{fedoseev16}. \citet{barone15} proposed that formamide  could form due to the reaction between H$_2$CO and NH$_2$ in the gas phase. Recently \citet{coutens16} detected the mono-deuterated forms of formamide using the ALMA-PILS survey and reported a deuteration level of $\sim$ 2\% assuming a standard $^{12}$C/$^{13}$C ratio of 68. \citet{skouteris17} also calculated the rate coefficients of the reactions producing deuterated formamide. Consequently, a comparison of the deuteration of formamide and formaldehyde can help determine if this gas phase pathway (for formamide) is possible or if it has to be ruled out in this source. According to the respective rate coefficients determined for NH$_2$CHO and NH$_2$CDO by \citeauthor{skouteris17}, the HDCO/H$_2$CO ratio should be three times higher than the NH$_2$CDO/NH$_2$CHO ratio if this gas-phase reaction occurs. At similar spatial scales, the best estimate HDCO/H$_2$CO ratio is 6.5$\pm$1\,\%, i.e.\ approximately three times higher than the NH$_2$CDO/NH$_2$CHO ratio. Thus, a gas phase formation pathway for formamide cannot be ruled out. Since the expected deuteration levels for the case of a grain surface formation pathway has not been explored, none of the scenarios for formamide formation can be ruled out from this comparison at this point.

\section{Summary and Outlook}\label{sect:future}

Measuring the deuterium fractionation is an important tool in understanding the early chemical evolution of forming protostars. The evolving picture is that the gas phase deuteration of molecules in the inner warm regions, representative of the bulk ice, seems to be lower than previously thought based on single dish observations. This highlights that more high-sensitivity and -resolution radio interferometric observations are needed to probe this gas reservoir and unlock the chemical history and future of deeply embedded low-mass protostars.

In this study we have detected several lines of various isotopologues of formaldehyde toward the deeply embedded low-mass protostar IRAS~16293$-$2422~B as part of the ALMA-PILS survey. Both H\down{2}C\up{17}O and D\down{2}\up{13}CO are detected for the first time in the ISM. Many of the lines are optically thick due to the high densities probed. The determined excitation temperature for HDCO is $106\pm13$~K, similar to several other molecules in the same survey, and previous measurements. 

Assuming the same excitation temperature constrained from HDCO for all the forms of formaldehyde to determine the column densities shows that many of the transitions are optically thick and abundances high. The detected forms of formaldehyde are: H\down{2}CO, H\down{2}\up{13}CO, H\down{2}C\up{17}O, H\down{2}C\up{18}O, HDCO, HDC\up{18}O, D\down{2}CO and D\down{2}\up{13}CO. The large number of lines covered by the unbiased spectral survey makes it possible to constrain the column density by focusing on the optically thin, unblended lines.

The measured HDCO/H\down{2}CO ratio is 6.5$\pm$1\%. This is significantly lower than what was previously estimated using single dish telescopes (14--33\%). This is also true for doubly deuterated formaldehyde, where the D\down{2}CO/HDCO ratio is 12.8\err{3.3}{4.1}\%, and D\down{2}CO/H\down{2}CO ratio is 0.6$\pm$0.1\%. The lines observed with single dish telescopes in previous studies also covered by the ALMA-PILS survey are significantly optically thick. The source size is better constrained here by spatially resolving the emitting region. These effects could be the reason for the different ratio(s).

These levels of deuterium fractionation are in line with formaldehyde forming on the grains once CO has frozen onto the surface as soon as the temperature drops low enough. The astrochemical model of \citet{taquet14} can approximately reproduce the deuterium fractionation observed in formaldehyde with a grain surface formation pathway dominating the production. However, the role of gas phase formation routes is still not clear. 

The effect seen for water  \citep{persson13,coutens12,furuya16}, where D\down{2}O/HDO is significantly higher than the HDO/H\down{2}O ratio is only minor for formaldehyde due to the even later start of formation toward the end of the prestellar core phase in comparison to water. This also implies higher absolute deuterium fractionation (D/H$_\mathrm{H_2CO}\sim$3\%, D/H$_\mathrm{H_2O}\sim$0.1\%), due to the more favorable conditions (low o/p-ratio, low $T$, higher accretion of D atoms onto grains). The deuterium fractionation presented for formaldehyde is similar to that of other molecules with similar complexity such as methanol \citep[CH\down{3}OH][]{jorgensen17} and formamide \citep[NH\down{2}CHO,][]{coutens16}, at around 2--3\%. This formation-time dependence on deuteration implies that more complex molecules which form even later should have even higher levels of deuterium fractionation.

The \up{12}C/\up{13}C and \up{18}O/\up{17}O ratios in formaldehyde are consistent with the values measured for the surrounding ISM at the relevant galactocentric distances. This could indicate a relatively common formation environment in the Galaxy. However, the \up{16}O/\up{18}O ratio is slightly higher than the local ISM value thus it is not possible to draw any strong conclusions based on this.

Future high resolution interferometric observations constraining the deuterium fractionation in formaldehyde of other sources, could shed light on any possible dependence on environment and mass. In addition to this, multi transition single dish observations unambiguously characterizing the cold gas would improve the global picture of formaldehyde deuterium fractionation.

\begin{acknowledgements}
  This research made use of Astropy and Astroquery, community-developed Python packages 
  for Astronomy \citep{astropy13,astroquery17}.
  MVP postdoctoral position is funded by the ERC consolidator grant 614264.
  JKJ acknowledges support from the European Research Council (ERC) under the European  Union’s 
  Horizon 2020 research and innovation programme (grant agreement No 646908) 
  through ERC Consolidator Grant “S4F”. Research at Centre for Star and Planet 
  Formation is funded by the Danish National Research Foundation. 
  AC postdoctoral grant is funded by the ERC Starting Grant 3DICE (grant agreement 336474).
  VT has received funding from the European Union's Horizon 2020 research and innovation programme under the Marie Sk\l{}odowska-Curie grant agreement n. 664931. 
  EvD acknowledges EU A-ERC grant 291141 CHEMPLAN and a KNAW professorship prize.
  This paper makes use of the following ALMA data: ADS/JAO.ALMA\#2013.1.00278.S. 
  ALMA is a partnership of ESO (representing its member states), NSF (USA) and 
  NINS (Japan), together with NRC (Canada) and NSC and ASIAA (Taiwan), in 
  cooperation with the Republic of Chile. The Joint ALMA Observatory is operated 
  by ESO, AUI/NRAO and NAOJ.
\end{acknowledgements}

\bibliographystyle{aa} 
\bibliography{bibfile}

\clearpage

\begin{appendix}

\section{Formaldehyde lines used in the analysis}\label{app:table_lines}
\begin{table*}
	\caption{Lines detected in the ALMA-PILS spectrum of IRAS 16293$-$2422. The stars in the left-most column indicates the lines that are blended with each other. The square in the right-most column identifies the lines used in constraining the column density (and for HDCO also the temperature).}\label{tab:lines}
	\centering
	\begin{tabular}{lccclrc}
		\hline\hline
		Species & Q$_1$ & Q$_2$ & $\nu$ & A$_{ij}$ & E$_{u}$ & Fit? \\
		& $J_{K_A K_B}$ &  & $\mathrm{GHz}$ & $\log_{10}$()  & $\mathrm{K}$  &   \\
		\hline
	H\down{2}CO 			& 5$_{1,5}$ & 4$_{1,4}$ & 351.769 & -2.9201 & 62.5 &   \\
							& 8$_{2,6}$ & 9$_{0,9}$ & 361.968 & -5.9856 & 173.5 & $\blacksquare$ \\
							& 5$_{0,5}$ & 4$_{0,4}$ & 362.736 & -2.8626 & 52.3  &   \\
	H\down{2}C\up{17}O 		& 5$_{1,5}$ & 4$_{1,4}$ & 343.333 & -2.9518 & 61.3  &  \\
							& 5$_{0,5}$ & 4$_{0,4}$ & 353.820 & -2.8950 & 51.0  &  $\blacksquare$  \\
							& 5$_{2,4}$ & 4$_{2,3}$ & 354.908 & -2.9665 & 98.4  & $\blacksquare$ \\
	\multicolumn{1}{r}{*}  	& 5$_{4,2}$ & 4$_{4,1}$ & 355.042 & -3.3340 & 240.1  &  \\
	\multicolumn{1}{r}{*} 	& 5$_{4,1}$ & 4$_{4,0}$ & 355.042 & -3.3340 & 240.1  &  \\
	\multicolumn{1}{r}{**} 	& 5$_{3,3}$ & 4$_{3,2}$ & 355.202 & -3.0835 & 157.5  &  \\
							& 5$_{3,2}$ & 4$_{3,1}$ & 355.214 & -3.0834 & 157.5  & $\blacksquare$ \\
							& 5$_{2,3}$ & 4$_{2,2}$ & 356.187 & -2.9619 & 98.5  & $\blacksquare$ \\
	H\down{2}C\up{18}O & 5$_{1,5}$ & 4$_{1,4}$ & 335.816 & -2.9805 & 60.2 &  \\
					   & 5$_{0,5}$ & 4$_{0,4}$ & 345.881 & -2.9246 & 49.9 & \\
					   & 5$_{2,4}$ & 4$_{2,3}$ & 346.869 & -2.9963 & 97.4 & $\blacksquare$ \\
					   & 5$_{4,2}$ & 4$_{4,1}$ & 346.984 & -3.3639 & 239.6 & $\blacksquare$ \\
					   & 5$_{4,1}$ & 4$_{4,0}$ & 346.984 & -3.3639 & 239.6 & $\blacksquare$ \\
					   & 5$_{3,3}$ & 4$_{3,2}$ & 347.134 & -3.1134 & 156.7 &  \\
					   & 5$_{3,2}$ & 4$_{3,1}$ & 347.144 & -3.1134 & 156.7 &  \\
					   & 5$_{2,3}$ & 4$_{2,2}$ & 348.032 & -2.9920 & 97.5 & $\blacksquare$ \\
					   & 5$_{1,4}$ & 4$_{1,3}$ & 357.741 & -2.8982 & 63.4 &  \\
	H\down{2}\up{13}CO 		& 5$_{1,5}$ & 4$_{1,4}$ & 343.326 & -2.9517 & 61.3 &  \\
							& 5$_{0,5}$ & 4$_{0,4}$ & 353.812 & -2.8949 & 51.0  &  \\
							& 5$_{2,4}$ & 4$_{2,3}$ & 354.899 & -2.9664 & 98.4 &  \\
							& 5$_{4,2}$ & 4$_{4,1}$ & 355.029 & -3.3339 & 240.1  &  \\
							& 5$_{4,1}$ & 4$_{4,0}$ & 355.029 & -3.3339 & 240.1 &  \\
	\multicolumn{1}{r}{*}  	& 12$_{1,11}$ & 12$_{1,12}$ & 355.042 & -4.7277 & 285.4  & $\blacksquare$ \\
							& 5$_{3,3}$ & 4$_{3,2}$ & 355.191 & -3.0835 & 157.5   &\\
	\multicolumn{1}{r}{**}  & 5$_{3,2}$ & 4$_{3,1}$ & 355.203 & -3.0834 & 157.5 &  \\
							& 5$_{2,3}$ & 4$_{2,2}$ & 356.176 & -2.9618 & 98.5 &\\
	HDCO &  5$_{1,4}$ & 4$_{1,3}$ & 335.097 & -2.9841 & 56.2  &  \\
		 & 17$_{2,15}$ & 17$_{2,16}$ & 340.801 & -4.5839 & 517.6  & $\blacksquare$ \\
		 &  3$_{1,3}$ & 2$_{0,2}$ & 343.284 & -5.2876 & 25.8  &  \\
		 & 11$_{1,10}$ & 11$_{1,11}$ & 346.739 & -4.6725 & 219.4  &  \\
		 &  4$_{2,2}$ & 5$_{0,5}$ & 347.286 & -6.1800 & 62.9 & $\blacksquare$ \\
		 & 10$_{1,9}$ & 10$_{0,10}$ & 355.075 & -5.2163 & 184.3  & $\blacksquare$ \\
	HDC\up{18}O & 6$_{1,6}$ & 5$_{1,5}$ & 353.098 & -2.9039 & 67.4 & $\blacksquare$ \\
	D\down{2}CO & 6$_{1,6}$ & 5$_{1,5}$ & 330.674 & -2.9892 & 61.1 &  \\
	 & 6 $_{0,6}$ & 5$_{0,5}$ & 342.522 & -2.9331 & 58.1 &   \\
	 & 15$_{2,13}$ & 15$_{2,14}$ & 345.075 & -4.4700 & 370.5 & $\blacksquare$ \\
	 & 6 $_{2,5}$  & 5 $_{2,4}$ & 349.631 & -2.9553 & 80.4 &   \\
	 & 6 $_{5,2}$  & 5 $_{5,1}$ & 351.196 & -3.4129 & 193.7 & $\blacksquare$ \\
	 & 6 $_{5,1}$  & 5 $_{5,0}$ & 351.196 & -3.4129 & 193.7 &  $\blacksquare$ \\
	 & 6 $_{4,3}$  & 5 $_{4,2}$ & 351.487 & -3.1522 & 145.2 &  \\
	 & 6 $_{4,2}$  & 5 $_{4,1}$ & 351.492 & -3.1522 & 145.2 &   \\
	 & 6 $_{3,4}$  & 5 $_{3,3}$ & 351.894 & -3.0204 & 107.6  &  \\
	 & 6 $_{3,3}$  & 5 $_{3,2}$ & 352.244 & -3.0192 & 107.6  &  \\
	 & 6 $_{2,4}$  & 5 $_{2,3}$ & 357.871 & -2.9248 & 81.2 &   \\
	D\down{2}\up{13}CO & 6$_{0,6}$ & 5$_{0,5}$ & 337.552 & -2.9461 & 57.2  &  \\
	 & 6$_{2,5}$ & 5$_{2,4}$ & 344.225 & -2.9696 & 79.6  & $\blacksquare$ \\
	 & 6$_{4,3}$ & 5$_{4,2}$ & 345.957 & -3.1670 & 144.7  & $\blacksquare$  \\
	 & 6$_{4,2}$ & 5$_{4,1}$ & 345.961 & -3.1670 & 144.7  & $\blacksquare$ \\
	 & 6$_{3,4}$ & 5$_{3,3}$ & 346.346 & -3.0352 & 106.9  &  \\
	 & 6$_{3,3}$ & 5$_{3,2}$ & 346.662 & -3.0340 & 106.9 &   \\
	 & 6$_{2,4}$ & 5$_{2,3}$ & 351.961 & -2.9406 & 80.3 & $\blacksquare$ \\
	 & 6$_{1,5}$ & 5$_{1,4}$ & 360.989 & -2.8691 & 66.2  & $\blacksquare$ \\
		\hline
	\end{tabular}
\end{table*}

\begin{table*}
	\caption{Lines not detected but used to constrain the column densities. }\label{tab:lines_ulims}
	\centering
	\begin{tabular}{lccclr}
		\hline\hline
		Species & Q$_1$ & Q$_2$ & $\nu$ & A$_{ij}$ & E$_{u}$ \\
		& $J_{K_A K_B}$ &  & $\mathrm{GHz}$ & $\log_{10}$()  & $\mathrm{K}$ \\
		\hline
H\down{2}CO 		& 28$_{3,25}$ & 28$_{3,26}$ & 336.14 & -4.6821 & 1542.1 \\
H\down{2}C\up{17}O 	& 18$_{3,15}$ & 19$_{1,18}$ & 354.222 & -6.0022 & 691.4 \\
					& 12$_{1,11}$ & 12$_{1,12}$ & 355.087 & -4.7276 & 285.4 \\
H\down{2}C\up{18}O 	& 12$_{1,11}$ & 12$_{1,12}$ & 339.498 & -4.7872 & 279.1 \\
H\down{2}\up{13}CO 	& 29$_{3,26}$ & 29$_{3,27}$ & 349.48 & -4.6636 & 1606.4 \\
HDCO    			& 24$_{3,21}$ & 24$_{3,22}$ & 348.965 & -4.5321 & 1017.8 \\
HDC\up{18}O 		& 2$_{2,0}$ & 3$_{1,3}$ & 332.193 & -6.2815 & 41.0 \\
D\down{2}CO 		& 7$_{3,4}$ & 8$_{1,7}$ & 340.380 & -6.1185 & 127.4 \\
		\hline
	\end{tabular}
\end{table*}

\section{Observed spectrum of lines}\label{app:figs_lines}
\begin{figure*}[htp]
	\centering
		\includegraphics[width=0.9\linewidth]{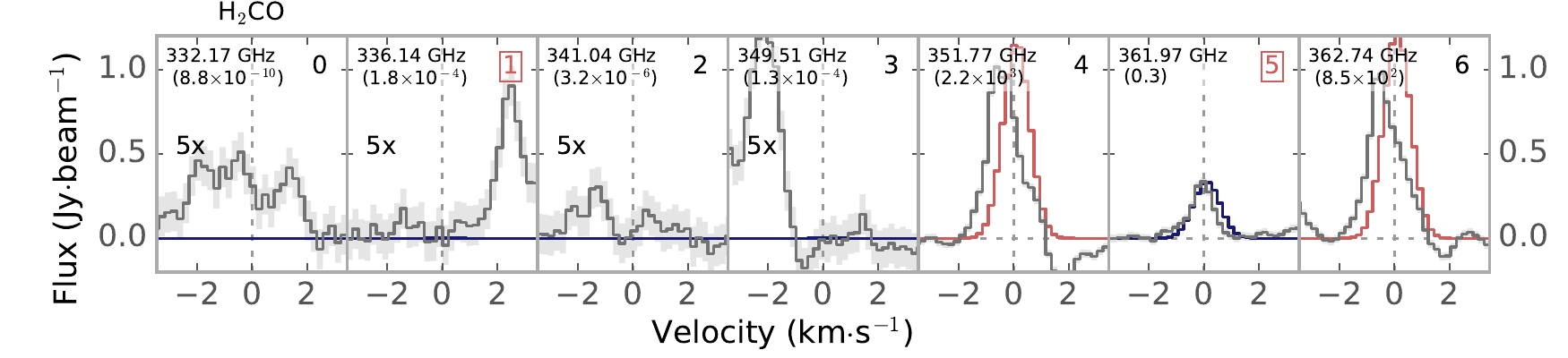}
	
		\includegraphics[width=0.9\linewidth]{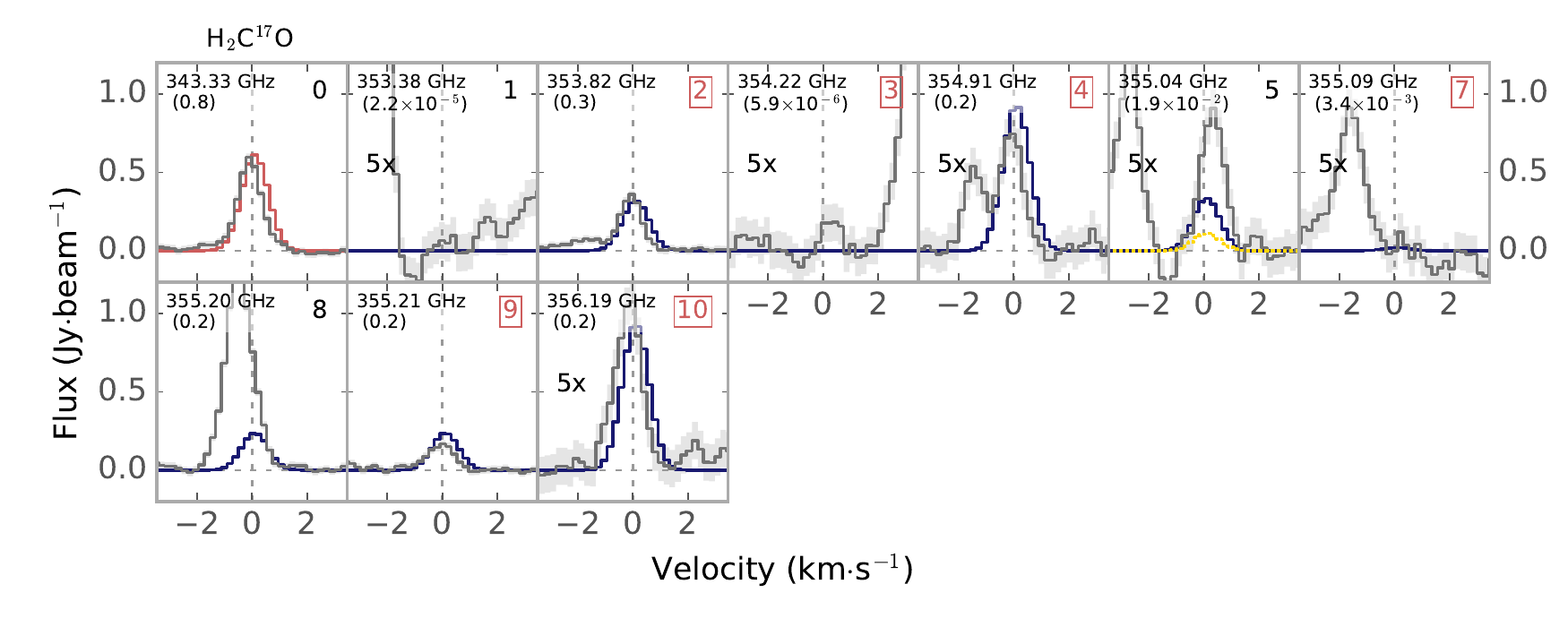}

		\includegraphics[width=0.9\linewidth]{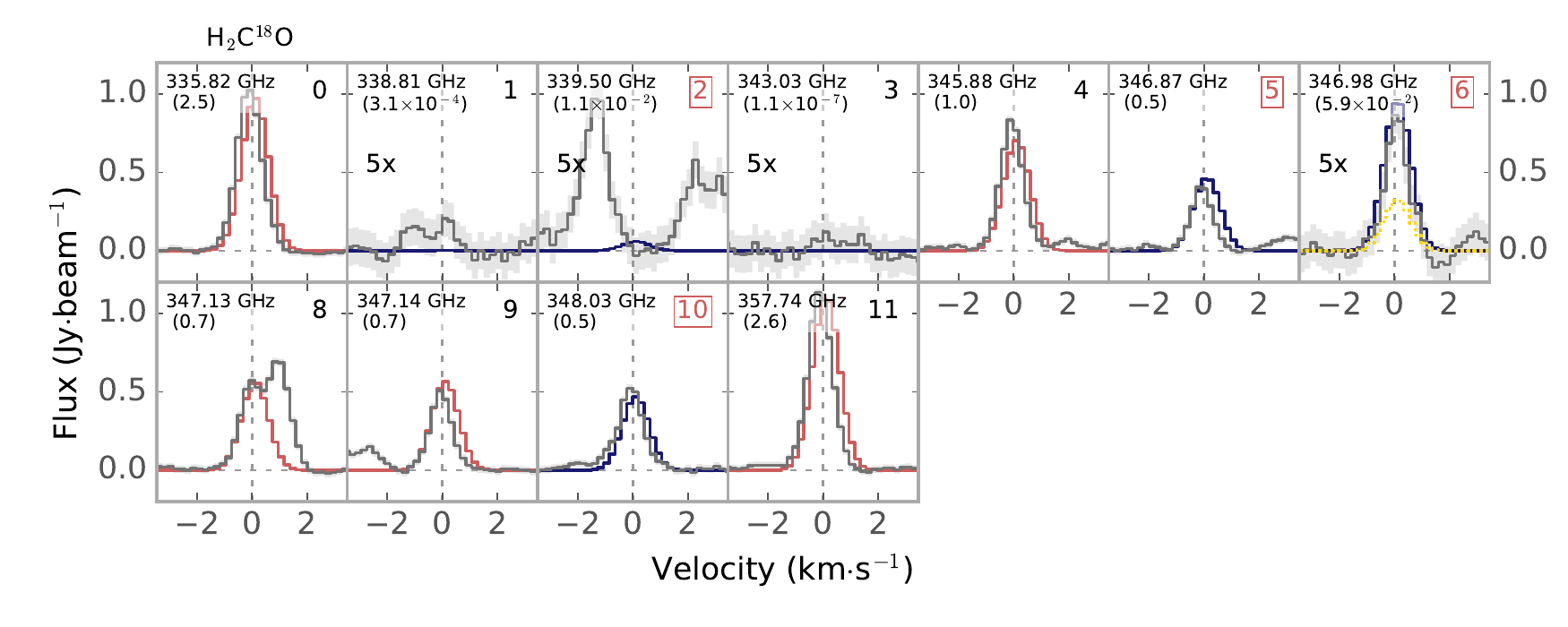}%
	
	\caption{All lines of (a) H\down{2}CO, (b) H\down{2}C\up{17}O, and (c) H\down{2}C\up{18}O with the best fit LTE model overplotted in blue. Synthetic spectra in red indicate optical depth beginning to play a significant role i.e.\ $\tau>0.6$. Numbers in top right corner with red boxes indicate lines used to fit the column density. Note that the two optically thick lines of H\down{2}CO are the ones with a significant amount of extended emission (covered in van der Wiel, in prep.). The number in parenthesis is the estimated optical depth. Lines in figure (b) panel 5 and 8 have stronger H\down{2}\up{13}CO lines blended (see panel 9 and 11 in \figref{fig:h213co_lte_spectrum}).}
	\label{fig:h2co_lte_spectrum}
\end{figure*}

	\begin{figure*}[ht]
		\centering
		\includegraphics[width=.9\linewidth]{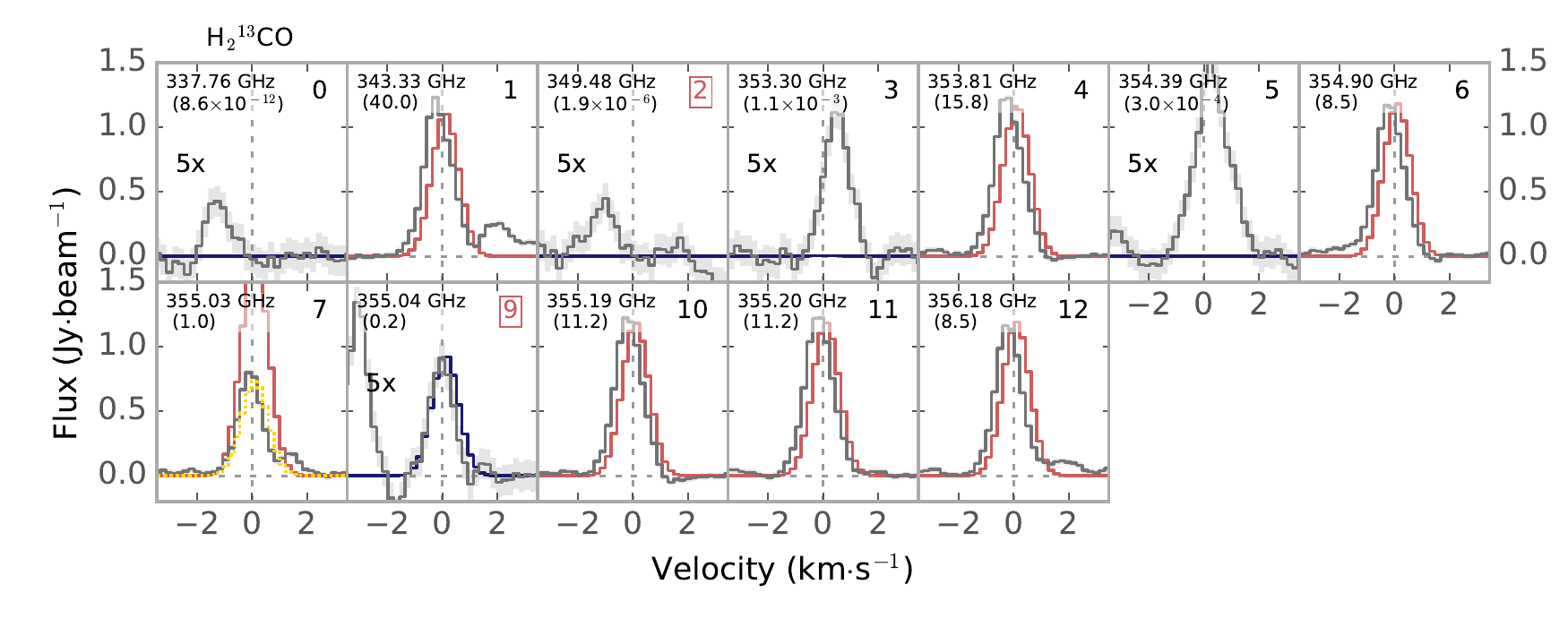}
		\caption{All lines of H\down{2}\up{13}CO with the best fit LTE model overplotted. Synthetic spectra in red indicate optical depth beginning to play a significant role i.e.\ $\tau>0.6$. Numbers in top right corner with red boxes indicate lines used to fit the column density.  The number in parenthesis is the estimated optical depth. The line in panels 9 and 11, at 355.04 and 355.20~GHz could be blended with significantly weaker H\down{2}C\up{17}O lines (see panel 5 and 8 in \figref{fig:h2co_lte_spectrum}).}
		\label{fig:h213co_lte_spectrum}
	\end{figure*}


	\begin{figure*}[ht]
		\centering
		\includegraphics[width=.9\linewidth]{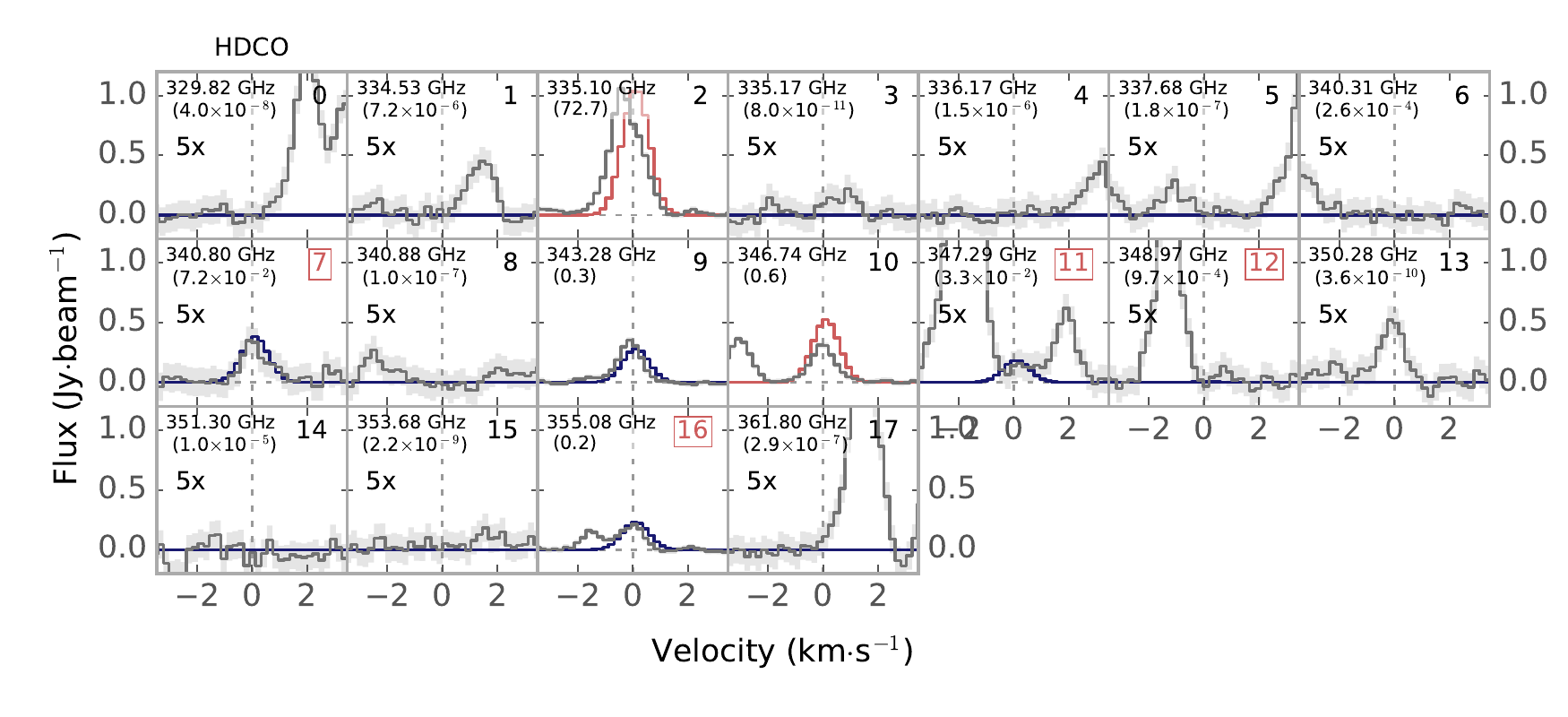}
		\caption{All the lines of HDCO with the best fit LTE model overplotted. Synthetic spectra in red indicate optical depth beginning to play a significant role i.e.\ $\tau>0.6$. Numbers in top right corner with red boxes indicate lines used to fit the column density. The number in parenthesis is the estimated optical depth.}
		\label{fig:hdco_lte_spectrum}
	\end{figure*}
	
	\begin{figure*}[ht]
		\centering
		\includegraphics[width=.9\linewidth]{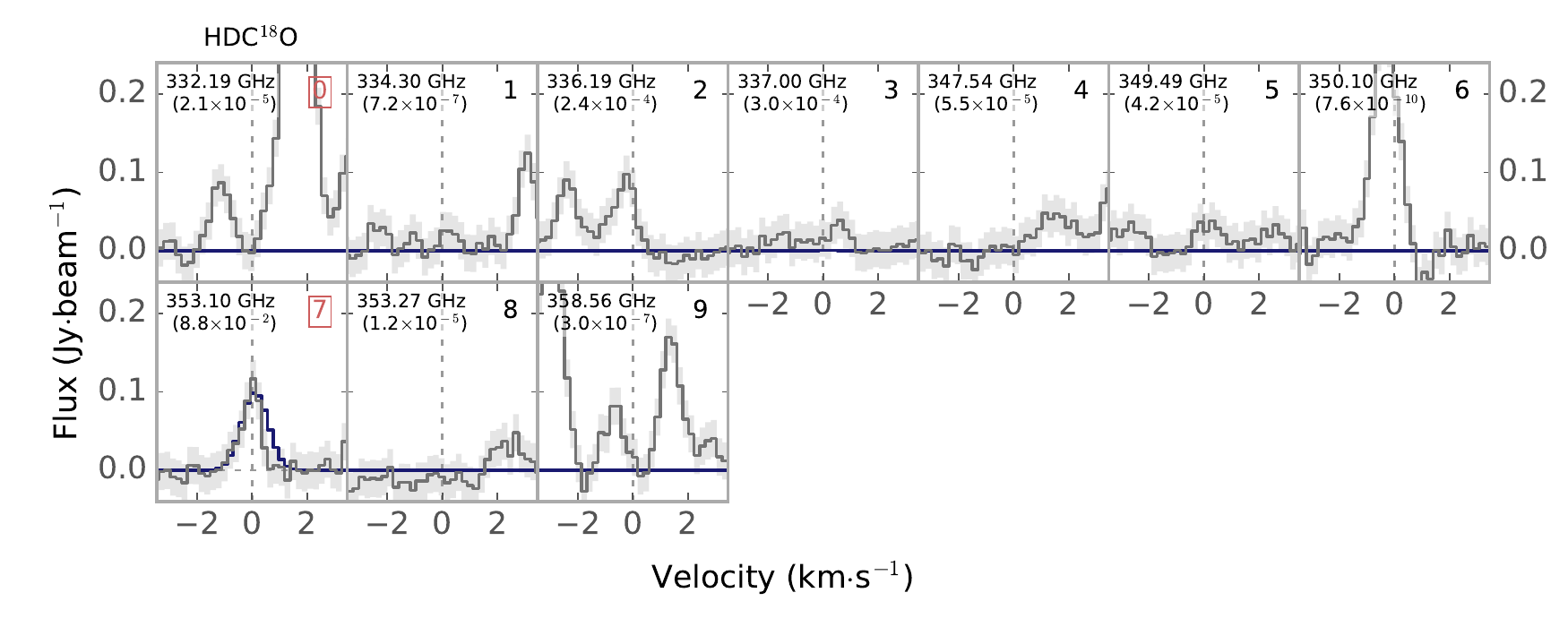}
		\caption{All the lines of HDC\up{18}O with the best fit LTE model overplotted. Synthetic spectra in red indicate optical depth beginning to play a significant role i.e.\ $\tau>0.6$. Numbers in top right corner with red boxes indicate lines used to fit the column density. The number in parenthesis is the estimated optical depth. The emission line at 350.10~GHz is probably methanol.}
		\label{fig:hdc18o_lte_spectrum}
	\end{figure*}


	\begin{figure*}[ht]
		\centering
		\includegraphics[width=.9\linewidth]{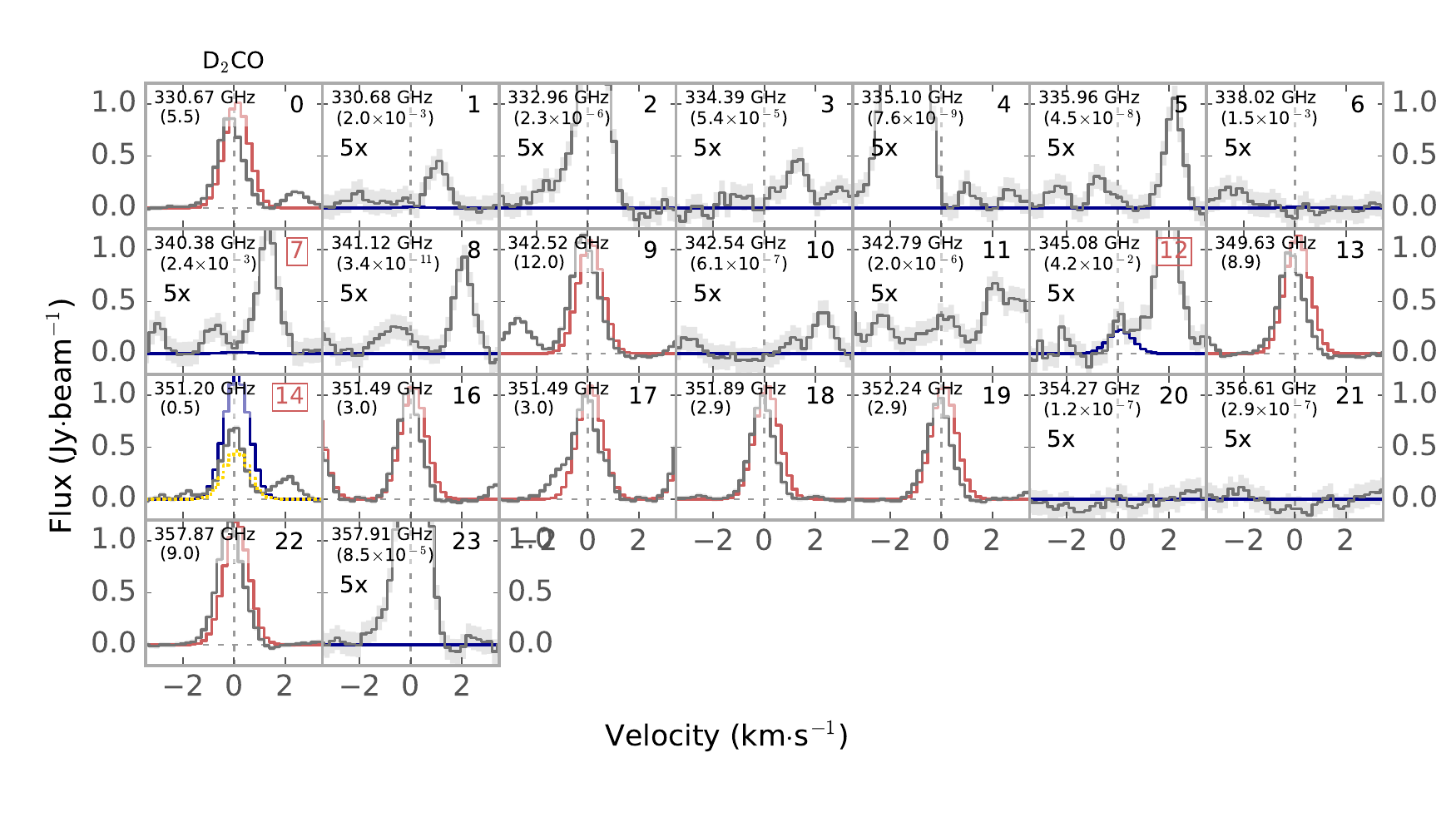}
		\caption{All the lines of D\down{2}CO with the best fit LTE model overplotted. Synthetic spectra in red indicate optical depth beginning to play a significant role i.e.\ $\tau>0.6$. Numbers in top right corner with red boxes indicate lines used to fit the column density. The emission line at 332.96~GHz is possibly glycolaldehyde and ethylene glycol, and at 357.91~GHz methyl formate.}
		\label{fig:d2co_lte_spectrum}
	\end{figure*}
	
	\begin{figure*}[ht]
		\centering
		\includegraphics[width=.9\linewidth]{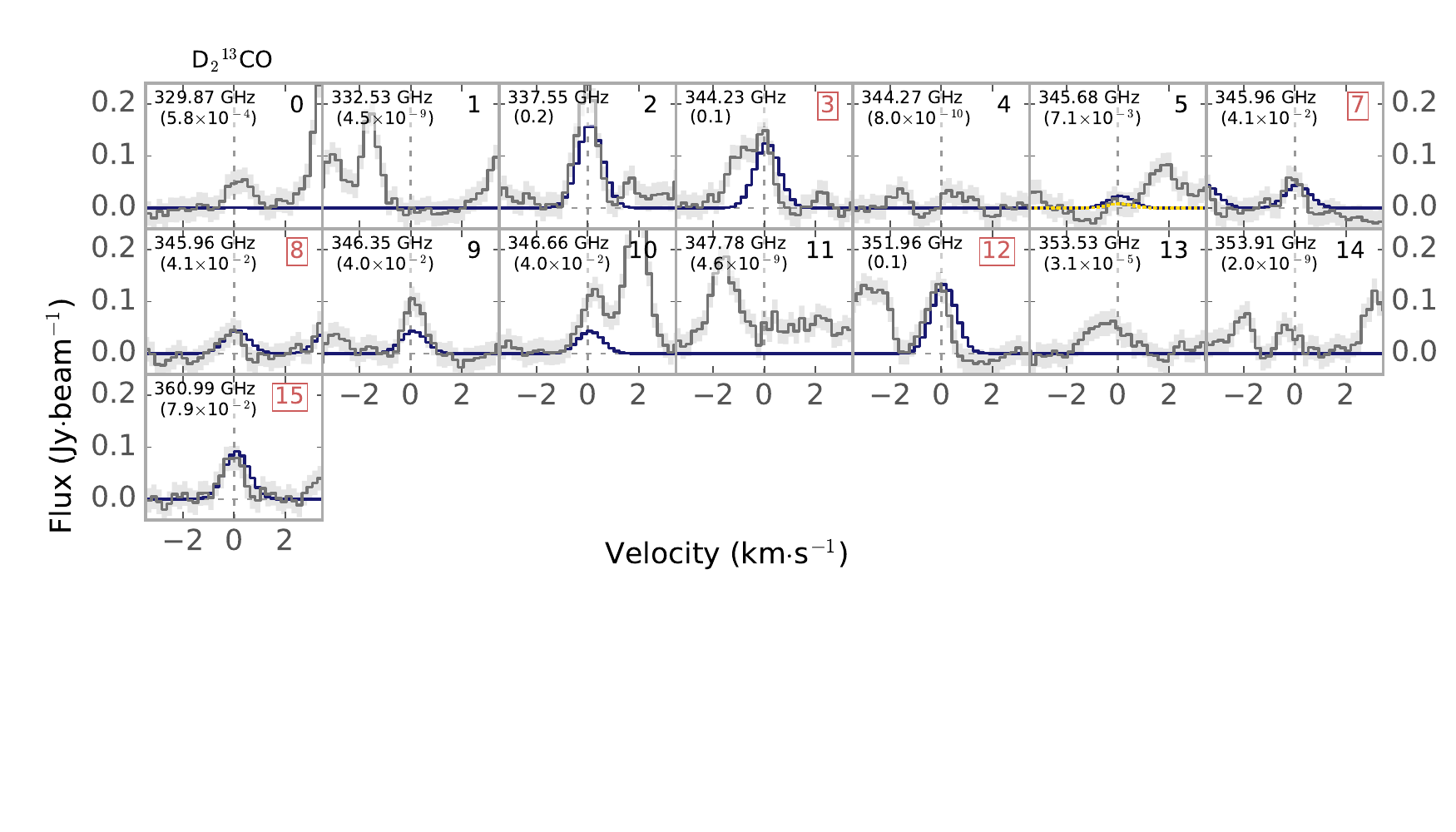}
		\caption{All the lines of D\down{2}\up{13}CO with the best fit LTE model overplotted. Synthetic spectra in red indicate optical depth beginning to play a significant role i.e.\ $\tau>0.6$. Numbers in top right corner with red boxes indicate lines used to fit the column density. The number in parenthesis is the estimated optical depth.}
		\label{fig:d213co_lte_spectrum}
	\end{figure*}

 \end{appendix}

\end{document}